\documentclass[aps,pra,twocolumn,showpacs,floatfix,superscriptaddress]{revtex4-1}
\usepackage{amsmath,amssymb,amsfonts,bbm,graphicx,hyperref,times}
\usepackage{mathptmx}
\usepackage{color}

\def\Tr{\hbox{Tr}}

\begin{document}
\title{Squeezing as a resource to counteract phase diffusion in optical phase estimation}
\author{G. Carrara}
\affiliation{Institut f\"ur Theoretische Physik III, Heinrich-Heine-Universit\"at D\"usseldorf, Universit\"atsstr. 1, D-40225 D\"usseldorf, Germany}
\author{M. G. Genoni}
\author{S. Cialdi}
\author{M. G. A. Paris}
\author{S. Olivares}\email{stefano.olivares@fisica.unimi.it}
\affiliation{Dipartimento di Fisica ``Aldo Pontremoli'', Universit\`a degli Studi di Milano, I-20133 Milano, Italy}
\affiliation{INFN - Sezione di Milano, I-20133 Milano, Italy}
\begin{abstract}
We address a phase estimation scheme using Gaussian states in the presence of 
non-Gaussian phase noise. At variance with previous analysis, we analyze situations in which 
the noise occurs before encoding phase information. In particular, we study 
how squeezing may be profitably used before or after phase diffusion.
Our results show that squeezing the probe after the noise greatly enhances the 
sensitivity of the estimation scheme, as witnessed by the increase of the quantum Fisher information. We then consider a realistic setup where homodyne detection is employed 
at the measurement stage, and address its optimality as well as its performance 
in the two different scenarios.
\end{abstract}
\maketitle
\section{Introduction}\label{0}
The problem of estimating a parameter, when a direct measurement of it is not feasible, is of crucial importance, in particular in quantum mechanics where not all the physical parameters correspond to observables. In this respect the estimation of an optical phase has been extensively studied in literature and both theoretical attempts to define a Hermitian operator \cite{PHASE1,PHASE2,PHASE3} and practical methods to estimate an optical phase \cite{PHASEMEASURE1,PHASEMEASURE2,PHASEMEASURE3,PHASEMEASURE4,PHASEMEASURE5,PHASEMEASURE6,PHASEMEASURE7}  have been proposed, mostly involving optical interferometry and homodyne detection \cite{FI1,FI2,FI3}. The problem of measuring an optical phase is particularly relevant in quantum communication, due to the possibility of encoding and transmitting information through a phase shift on a quantum state, with all the advantages of using a quantum state for communication applications \cite{COMMUNICATION1,COMMUNICATION2}. Moreover, it is also of interest in quantum sensing, as the change in the optical phase may be also due, for instance, to the specific properties of biological samples \cite{Taylor2016}.
\par
In the past, different phase estimation scheme have been analyzed, in particular using pure probe Gaussian states \cite{PURE} and the advantages of introducing the squeezing has been highlighted. As a matter of fact, the squeezing has been shown to provide advantages in a wide range of applications \cite{SQADV0,SQADV00,SQADV000,SQADV1,SQADV2,SQADV3,ul15} and introducing the squeezing in a real optical system has been shown to be experimentally achievable with different methods, in particular using Optical Parametric Oscillators (OPO) \cite{SQUEEZING1}, involving non linear interactions in crystals. 
\par
In a realistic scenario, it is sometimes fundamental to include in the phase estimation scheme a model of the most detrimental type of noise, that is a phase-diffusive one. So far, phase noise has been investigated in different systems, ranging from qubits \cite{QUBITPD1,QUBITPD2} to condensate systems \cite{OTHERPD1,OTHERPD2} to Bose-Josephson junctions \cite{OTHERPD3}, and most important in Gaussian optical states for, which are the focus of this paper, for both phase estimation \cite{PRL,PRA} and quantum communication \cite{Comm01,Comm02,Comm03,Comm04,Comm05}. There, phase-diffusion typically describes the noise affecting the propagation of quantum light thorugh optical fibers \cite{IP2008}. However, the effects of the phase noise have been studied only at the detection stage \cite{NOISE1,NOISE2,NOISE3,NOISE4,NOISE5} or when this noise occurs between the encoding of the information and the detection \cite{PRL,PRA}, i.e. in the transmission of the encoded state.
\par
\begin{figure}[tb!]
\includegraphics[width=\columnwidth]{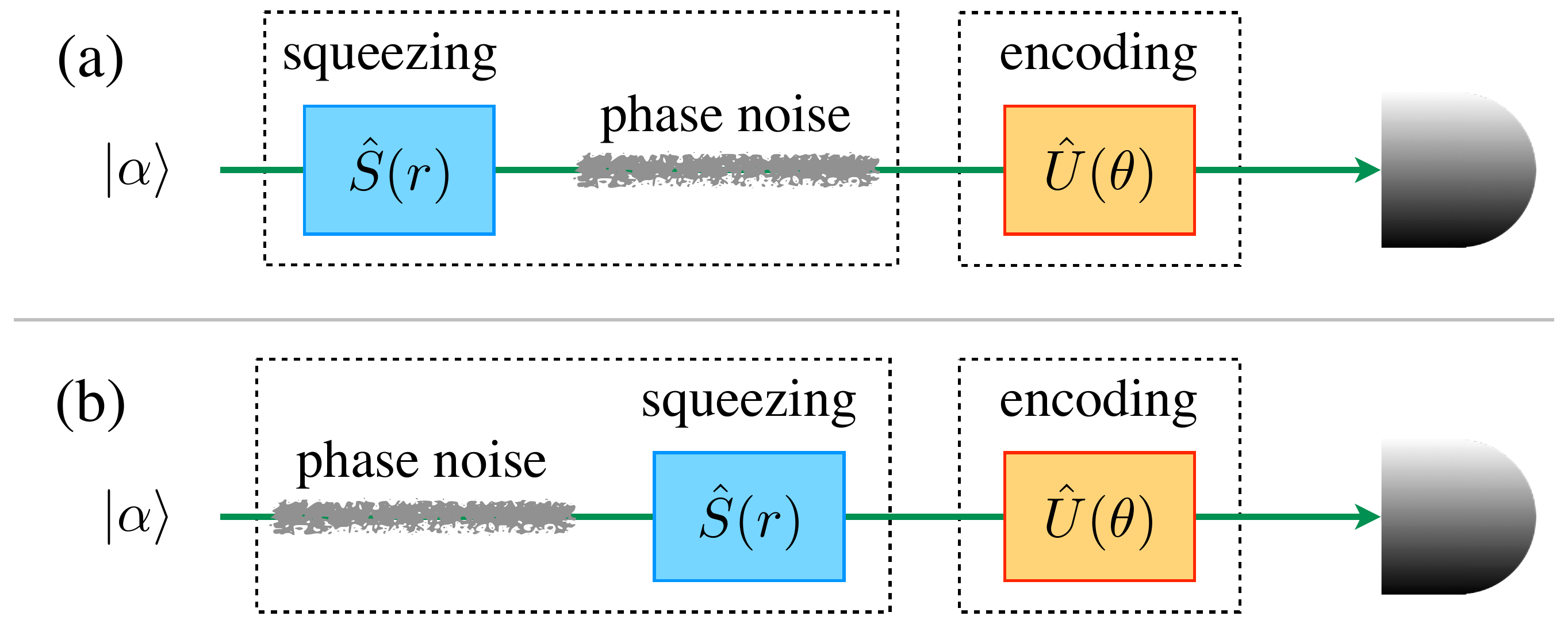}
\vspace{-0.6cm}
\caption{Sketches of the two different scenarios we shall analyze starting always from a coherent state $|\alpha \rangle$: (a) the squeezing $\hat{S}(r)$ is introduced {\it before} the phase noise and (b) the squeezing is applied {\it after} the phase noise. It is worth stressing that, differently from the previous works, the phase noise occurs always before the encoding of the information, represented by the phase shift $\hat{U}(\theta)$.
\label{f:scenarios}}
\end{figure}

In a recent work \cite{CialdiPRL}, it was demonstrated that the squeezing operation be may used to ``squeeze'' the phase noise affecting an input coherent state, thus leading to possible advantages for application in quantum estimation and communication. In this paper, motivated by these results, we shall address and characterize a phase estimation scheme in which phase diffusion affects the generation stage of the protocol, i.e. before the encoding of the information.
To give an operational interpretation of this result and both characterize and quantify the squeezing as a resource for this phase-estimation problem, we shall investigate the two scenarios depicted in Fig.~\ref{f:scenarios}: the
 one in which the phase noise occurs after the squeezing operation and before encoding, the other in which the phase noise affects the ``coherent seed state'' before the squeezing. To this aim, we shall evaluate the ultimate limits to estimation precision posed by the quantum Cram\'er--Rao bound, and we shall also look for regimes where homodyne detection is the optimal measurement. Therefore, beside the fundamental aspects of our research, our results could be also relevant in a multi-node communication protocol, where light is transmitted along optical fibers, and at each node users may have to either decode or further encode information via the optical phase.

The paper is structured as following: in Section \ref{sI} we shall briefly address the quantum estimation theory. In Section \ref{II} we then characterize all the elements of the phase estimation scheme we consider and we show the results of a phase estimation scheme using pure Gaussian states. Finally, in Section \ref{III} we present the results of the extended numerical evaluation for the ultimate bounds to estimation precision obtained using phase-diffuse input states and homodyne detection, summarizing the conclusions in Section \ref{IV}.
\section{Quantum Estimation theory}\label{sI}
Estimating a parameter is often a necessary task, since several physical quantities cannot be directly measured. The theory behind the estimation of a parameter is well known and studied and, here, we briefly review the main elements we shall use throughout this work. Given a set of quantum states $\varrho_\lambda$, $\lambda$ being the parameter we seek to estimate, we suppose to perform a certain measurement, represented by a positive-operator-valued measurement 
$\{\hat{\Pi}_x\}$, obtaining a set of $M$ outcomes $\Xi = \{ x_1,...x_M\}$, which are then processed by means of a map $\hat{\phi}(\Xi)$, called \emph{estimator}, which in turn provides us with an estimated value of the parameter $\lambda$. The conditional probability of getting an outcome $x$ given the parameter $\lambda$ follows from the so-called \emph{Born rule}, namely
\begin{align}
p(x | \lambda) = \mbox{Tr}[\varrho_\lambda \hat{\Pi}_x]\,.
\end{align}
If the estimator is unbiased, its variance is bounded by the quantity \cite{INFO4}
\begin{eqnarray}\label{FI}
F(\lambda)=\int dx \frac{1}{p(x |\lambda)} \left( \frac{\partial p(x|\lambda)}{\partial \lambda} \right)
\end{eqnarray}
called \emph{Fisher Information} (FI), via the so-called \emph{Cram\'er-Rao bound}, that is 
\begin{align}
\mbox{Var}(\lambda) \geq \frac{1}{M F(\lambda)}\,,
\end{align}
where the factor $M^{-1}$ follows from the statistical scaling.
Therefore, given a measurement, the minimum uncertainty obtainable when estimating a parameter is related to the FI of the probability density $p(x|\lambda)$. However, one may question whether a more strict bound, independent of the measurement chosen, can be found: the answer is positive, and an ultimate bound to precision is obtainable. To this aim, one defines the \emph{Symmetric Logarithmic Derivative} via the Lyapunov equation $2\partial_\lambda \varrho_\lambda = \hat{L}_\lambda \varrho_\lambda+\varrho_\lambda \hat{L}_\lambda$
and uses this quantity to define the \emph{Quantum Fisher Information} (QFI) as \cite{Helstrom,INFO4}
\begin{align}
H(\lambda) =  \mbox{Tr}[ \varrho_\lambda \hat{L}_\lambda^2]\,.
\end{align}
The QFI poses a lower bound for the variance of any estimator and performing any measurement, given by the so-called \emph{Quantum Cram\'er--Rao bound} \cite{QFI1,QFI2,QFI3,QFI4}
\begin{align}
\mbox{Var}(\lambda) \geq \frac{1}{M H(\lambda)}\,.
\end{align}

In an estimation scheme the FI and QFI are of crucial importance: given a measurement, the FI provides us with the minimum variance achievable by any estimator and the QFI represents the ultimate quantum limit to precision, independent from the measurement. In our work we shall evaluate, for the chosen estimation scheme, explained in detail below, both the QFI and the FI relative to the commonly used homodyne detection, thus determining whether such measurement is optimal.

\section{Setting the scene: noiseless phase estimation scheme}
\label{II}
We consider an optical field described by the quadrature operators $\hat{X} =(\hat{a}+\hat{a}^\dag)/\sqrt{2}$ and $\hat{Y}=i (\hat{a}^\dag - \hat{a})/\sqrt{2}$, typically dubbed respectively as amplitude and phase quadratures, and defined in terms of the bosonic operators satisfying the commutation relation $[\hat{a}, \hat{a}^\dag] = \mathbbm{1}$. We shall now introduce and characterize step by step the phase estimation scheme in the noiseless scenario, while in the following Section we shall introduce the phase-diffusive noise.

\begin{figure}[tb!]
\centering
\includegraphics[ width=5cm, keepaspectratio]{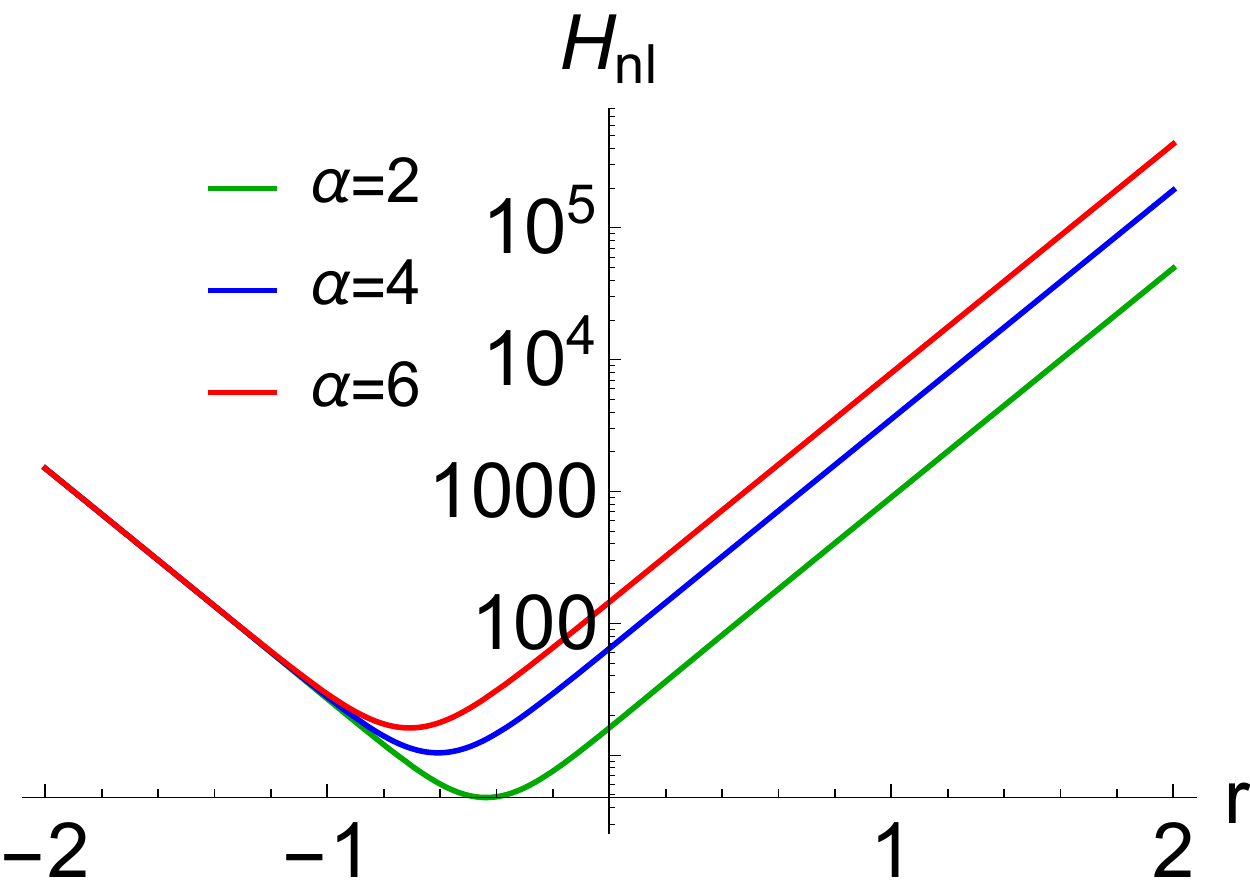}
\caption{Logarithmic plot of the QFI as a function of the squeezing parameter $r$ for different values of $\alpha$: in green (bottom line) $\alpha=2$, in blue (middle line) $\alpha=4$ and in red (top line) $\alpha=6$ (phase diffusion is fixed at $\sigma^2 = 0.01$). Two regimes can be identified in the plot: for positive values of the squeezing parameter, the second term of Eq.~(\ref{pureqfi}) dominates, thus leading to an exponential rise and to a dependance on the coherent amplitude. On the other hand, for negative values of $r$, the second term of Eq.~(\ref{pureqfi}) decays, and only the first term remains, leaving to a similar exponential rise, independent on the value of $\alpha$.}
\label{nopd}
\end{figure}

In the absence of noise  in the scenarios depicted in Fig.~\ref{f:scenarios}, the input Gaussian states before the encoding are just \emph{squeezed displaced states}, whose pure state reads
\begin{align}\label{state}
|\psi(r,\alpha)\rangle =  \hat{S}(r) \hat{D}(\alpha) | 0 \rangle
\end{align}
where $\hat{S}(r)=e^{\frac{1}{2}[r (\hat{a}^{\dagger})^2 - r^* \hat{a}^2]}$ and $\hat{D}(\alpha)=e^{\alpha \hat{a}^{\dagger} - \alpha^* \hat{a}}$ are the squeezing and the displacement operators, respectively \cite{COHERENT}. The information is now encoded through a phase shift, achieved with the \emph{phase shift operator}
\begin{align}
\hat{U}(\theta)=e^{-i\theta \hat{a}^{\dagger} \hat{a}}\,.
\label{eq:phaseshift}
\end{align}
The value $\theta$ of the phase shift is the target of our estimation. In general, if the information is encoded on a pure probe state  $|\psi_0\rangle$ through a unitary operator $\hat{U}(\theta)=e^{i\theta \hat{G}}$, being $\hat{G}$ the generator of the unitary group, a simple analytical expression of the QFI can be found, that is \cite{INFO4}
\begin{align}
H=4 \Delta^2 \hat{G}
\end{align}
where $\Delta^2 \hat{G} = \langle \psi_0 | \hat{G}^2 |\psi_0\rangle - \langle \psi_0 | \hat{G} |\psi_0\rangle^2$. It is worth noting that the QFI does not depend on the actual value of the parameter we seek to estimate. In our particular case, where the state is given by Eq.~(\ref{state}) and the generator of the unitary evolution is simply $\hat{G}= \hat{a}^{\dagger} \hat{a}$, the QFI in the noiseless case explicitly reads \cite{PURE,INFO4}
\begin{align}\label{pureqfi}
H_{\rm nl}=[\cosh(4r)-1]+4e^{4r}\alpha^2
\end{align}
where,{ in order to maximize its value and optimize the estimation precision, we assumed $\alpha, r \in  {\mathbbm R}$ (this choice will be adopted in the rest of the manuscript)}.
Therefore, the QFI depends on the coherent amplitude $\alpha$ and the squeezing parameter $r$, but not on the value of the phase shift $\theta$. In particular we are interested in determining whether the squeezing can have a meaningful impact of the phase estimation. In Fig.~\ref{nopd} we plot of the QFI as a function of $r$ for different values of the coherent amplitude. 
For our purposes the results are remarkable: a sensible improvement compared to the state with $r=0$ is indeed achieved with the introduction of a sufficient high squeezing, both for positive and negative values of the squeezing parameter. However, as we choose $\alpha \in \mathbbm{R}$, corresponding to a displaced vacuum state with a non-zero average value of the $\hat{X}$ quadrature, at fixed squeezing strength $|r|$, it is optimal to choose positive values 
\cite{PURE}. 

The final step to properly characterize our phase estimation scheme is the choice of the measurement. A natural choice in the presence of continuous-variable states is the so-called \emph{homodyne detection}, mathematically corresponding to projection on the eigenstates of either the amplitude or phase quadrature operators, or in general of any quadrature operator obtained by applying a phase-rotation. For pure Gaussian states as the one in Eq.~(\ref{state}) the homodyne probability distribution can be easily obtained from the vector of quadrature first moments and the corresponding covariance matrix \cite{QOPTICS4,GaussFOP, diffusione}. In particular, if we focus on the estimation of an infinitesimally small phase $\theta \approx 0$ (see Section~\ref{ss:discussion} for further details concerning this choice), one finds that the optimal quadrature to be measured is $\hat{Y}$, since $\alpha,r \in \mathbbm{R}$ \cite{PURE}. In the noiseless case we are able to evaluate an analytical expression for the FI in Eq.~(\ref{FI}) which, for our choice of estimated phase, reads
\begin{align}
F_{\rm nl}=4e^{4r}\alpha^2
\end{align}
We see that in the noiseless case this measure is indeed approximately optimal, that is the FI is approximately equal to the QFI for the relevant range of parameters investigated. In fact, for positive values of the squeezing parameter the first term of Eq.~(\ref{pureqfi}) is negligible compared to the second term, thus allowing us to write $H_{\rm nl} \approx 4e^{4r}\alpha^2 = F_{\rm nl}$. 

\section{Phase estimation in the presence of phase noise}
\label{III}

The scheme proposed so far shows the possibility of improving phase estimation thorough squeezing (if, of course, we choose $r > 0$). In real experimental setups, however, many detrimental and uncontrollable sources of noise are always present and must be taken into account. In particular, we shall consider the most detrimental source of noise for a phase estimation scheme, that is a phase diffusive noise. The phase diffusive noise is modelled as a random phase shift, with Gaussian distribution, applied to the state.

Let us consider a quantum state $\varrho$, affected by phase noise. Here we assume that the evolution is described by the master equation $\dot{\varrho} = \Gamma {\cal L}[\hat{a}^{\dagger} \hat{a}]$, where ${\cal L}[A]\varrho = 2 A \varrho A^{\dag} - A^{\dag}A \varrho - \varrho A^{\dag}A$, $\Gamma$ being the phase damping rate. Therefore, the evolved state can be thus written as \cite{PRL}
\begin{align}
\mathcal{E}_\sigma(\varrho) = \int_{\mathbbm R} d\psi \frac{e^{-\frac{\psi^2}{2 \sigma^2}}}{\sqrt{2\pi\,\sigma^2}}\,
\hat{U}(\psi) \varrho \hat{U}^\dagger(\psi)
\end{align}
where $\hat{U}(\psi)$ is the phase shift operator in Eq.~(\ref{eq:phaseshift}) and $\sigma^2 = 2 (\Gamma t)^2$ is the phase noise strength, $t$ being the evolution time. As we said, the difference with the previous works consist in the fact that the phase noise is introduced in the generation of the state, i.e. before the encoding of the information. As we have mentioned in the Introduction, since the squeezing operator and the phase noise do not commute, the presence of the phase noise in our scheme opens up the two different scenarios that we presented in Fig.~\ref{f:scenarios}: the first scenario, shown in (a), corresponds to the introduction of the squeezing operation before the phase noise, and thus to a family of input quantum states
\begin{align}
\varrho_{(a)} = \mathcal{E}_\sigma\left( |\psi(r,\alpha)\rangle\langle \psi(r,\alpha)|  \right) \,,  
\end{align}
where $\varrho(r,\alpha)$ represents the pure input states considered in Eq.~(\ref{state}). In the second scenario shown in Fig.~\ref{f:scenarios}~(b), the squeezing occurs after the phase noise, and thus corresponds to a family of input states,
\begin{align}
\varrho_{(b)} = \hat{S}(r) \mathcal{E}_\sigma(|\alpha\rangle\langle \alpha|) \hat{S}^\dagger(r) \,.
\end{align}
This scenario is, in principle, more interesting from a practical point of view, since it could be a suitable model to describe a real laser output, whose phase varies in time \cite{CialdiPRL}.

For both the scenarios a numerical evaluation of the QFI and the FI is needed to fully characterize the estimation scheme. It is moreover worth noting that in both cases, the introduction of the phase noise renders the state non-pure and non-Gaussian (in fact a Gaussian mixture of Gaussian states). Therefore for the QFI an analytical expression is not accessible anymore, thus compelling us to use approximate numerical methods. The method we shall use is based on a particular expression that shows the relationship between the QFI and the Bures distance (and consequently with the Fidelity between quantum states) \cite{BURES1,BURES2}:
\begin{align}
\label{eq:QFIFidelity}
H(\lambda)=8 \lim_{\delta\lambda \rightarrow 0}\frac{1-\mathcal{F}(\varrho_\lambda,\varrho_{\lambda+\delta\lambda})}{(\delta\lambda)^2}
\end{align}
where $\delta\lambda$ is an infinitesimal variation of the parameter $\lambda$, while $\mathcal{F}(\varrho,\tau) = \Tr[\sqrt{\sqrt\varrho \tau \sqrt\varrho}]$ represents the quantum fidelity between two states $\varrho$ and $\tau$. This form of the QFI can be used to evaluate numerically the QFI of a state, given its density matrix, also when an analytic form is not available, by choosing an appropriate small value of $\delta\lambda$. Of course, writing the exact density matrix for an infinite-dimensional full-rank state as the ones in Eq.~(\ref{state}) is not possible, but an approximate density matrix can be addressed, by writing the expressions of the field operators $\hat{a}$ and $\hat{a}^{\dagger}$ in the Fock basis and truncating the Hilbert space at a sufficiently large dimension $N_t$. For the range of parameters considered in this manuscript, we verified that the the choices $\delta\theta=0.005$ and $N_t=600$ led to consistent results.

\subsection{Comparison of ultimate limits for the two scenarios}
\begin{figure}[tb!]
\centering
{\includegraphics[width=0.48\columnwidth, keepaspectratio]{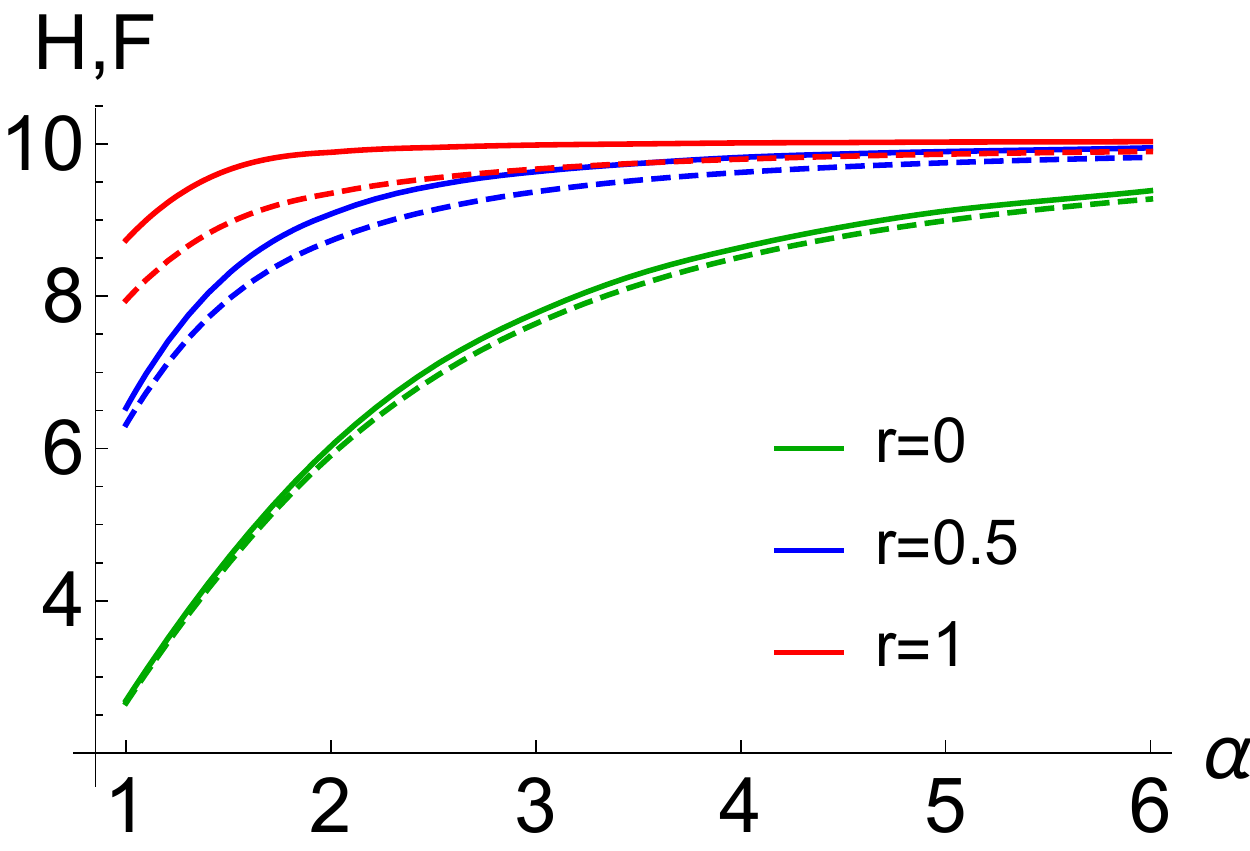}}
{\includegraphics[width=0.48\columnwidth, keepaspectratio]{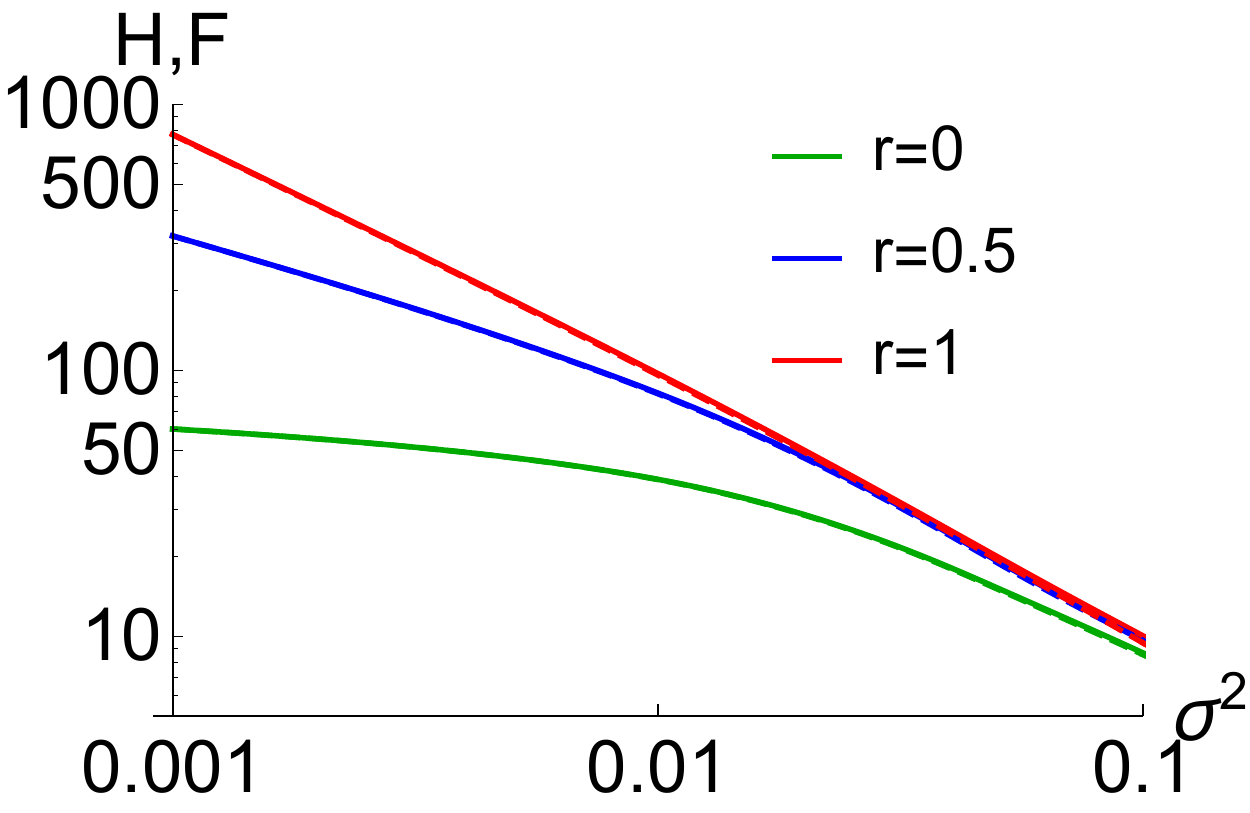}} \\
\vskip0.2cm
{\includegraphics[width=0.48\columnwidth, keepaspectratio]{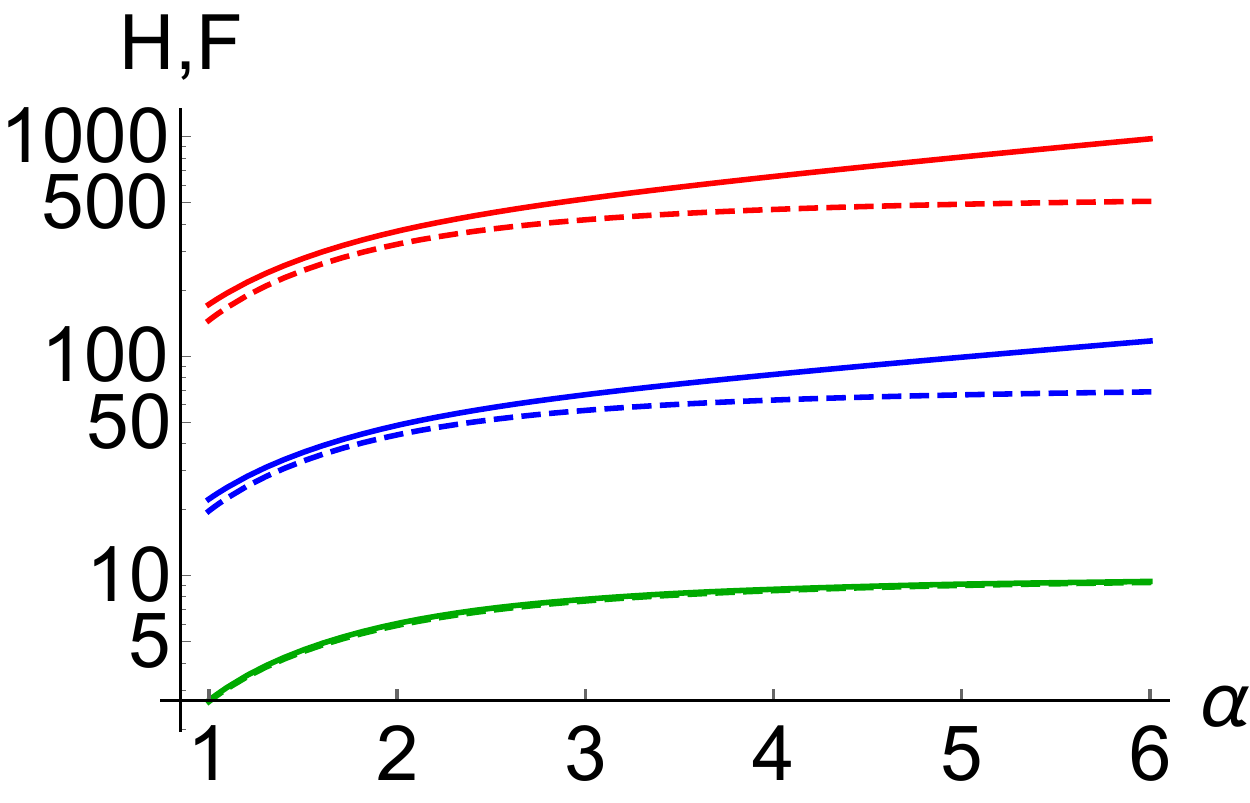}}
{\includegraphics[width=0.48\columnwidth, keepaspectratio]{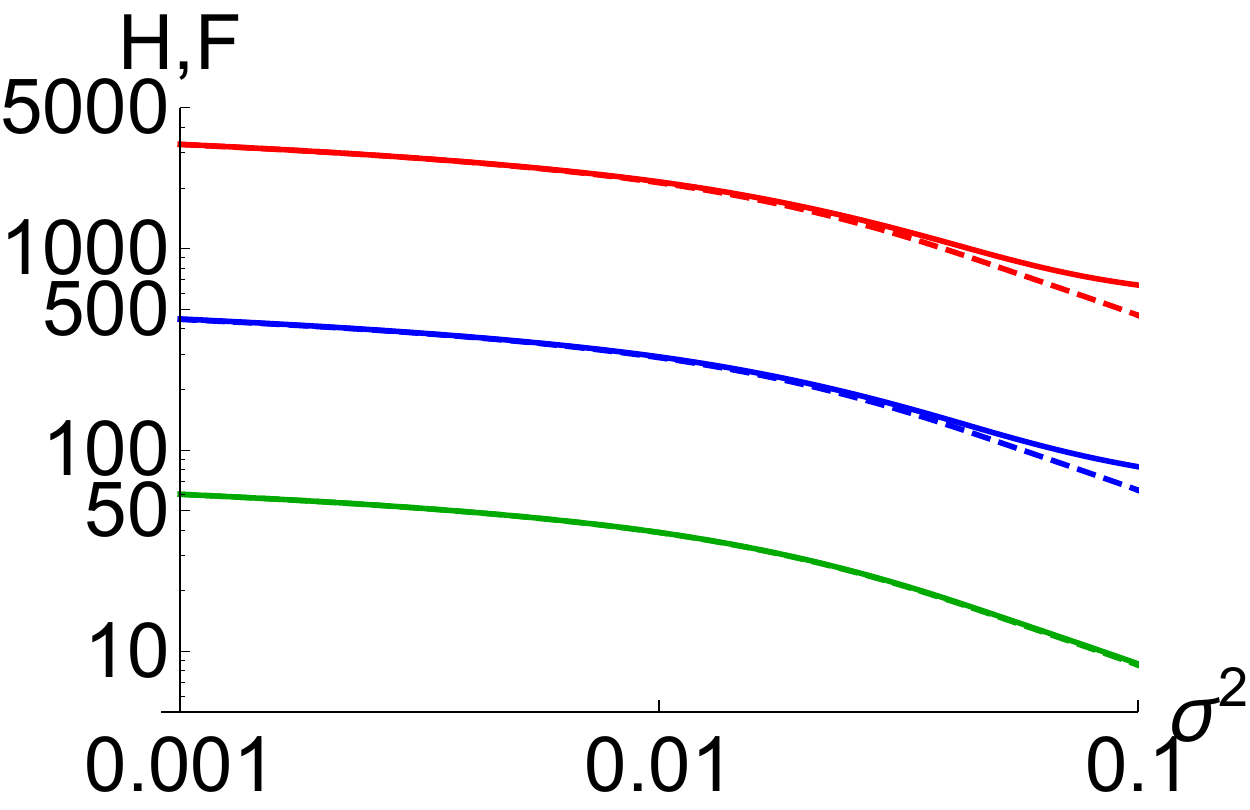}}
\caption{Plots of the QFI (solid lines) and FI corresponding to homodyne detection (dashed lines) for scenario (a) in the top panels and for scenario (b) in the bottom panels. QFI and FI are plotted as a function of the coherent state amplitude $\alpha$ (with $\sigma^2 = 0.1$) and of the dephasing noise $\sigma$ (with $\alpha=4$) respectively in the left and right panels, for different values of $r$: green (bottom lines), $r=0$; blue (middle lines), $r=0.5$; red (top lines), $r=1$.}
\label{QFIab_alpha_sigma}
\end{figure}

In the following we shall denote with $H_{(a)}$ and $H_{(b)}$ the QFI for the two families of input states, respectively $\varrho_{(a)}$ and $\varrho_{(b)}$, at fixed values of the different parameters $r$, $\alpha$ and $\sigma$. In the different panels of Fig. \ref{QFIab_alpha_sigma} we plot separately their behaviour as a function of either the coherent state amplitude $\alpha$ and phase-noise $\sigma$, for different values of (positive) squeezing $r$. As we expected in both cases we observe that the QFIs increases monotonically with $\alpha$ and $|r|$ and decreases monotonically with $\sigma$.

However we observe that in the scenario (a) in Fig.~\ref{f:scenarios}, where the squeezing is introduced before the phase noise, no relevant advantage can be obtained by the introduction of the squeezing, apart from low $\alpha$ and low noise regimes. In fact increasing either $\alpha$ or $\sigma$, leads to a saturation of the QFI independently on the value of $r$. This in particular allows us to conclude that increasing the coherent amplitude is a more feasible and experimentally achievable method to improve the phase estimation. The results for the scenario (b) in Fig.~\ref{f:scenarios} are remarkably different: by applying squeezing after the phase-noise one in fact obtains much higher values of QFI, and more importantly, this enhancement is maintained also when one increases the coherent state amplitude $\alpha$ or the noise parameter $\sigma$.
\begin{figure}[tb!]
\centering
{\includegraphics[width=0.48\columnwidth, keepaspectratio]{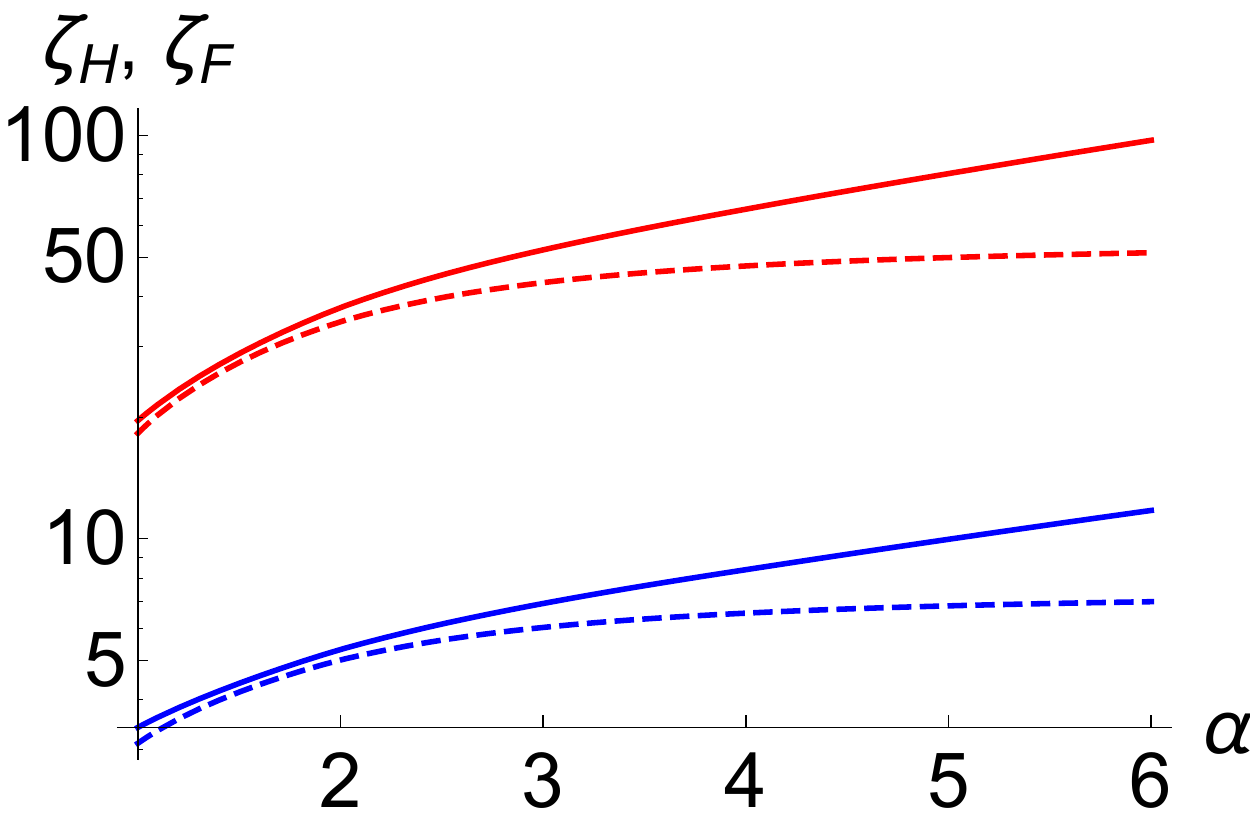}}
{\includegraphics[width=0.48\columnwidth, keepaspectratio]{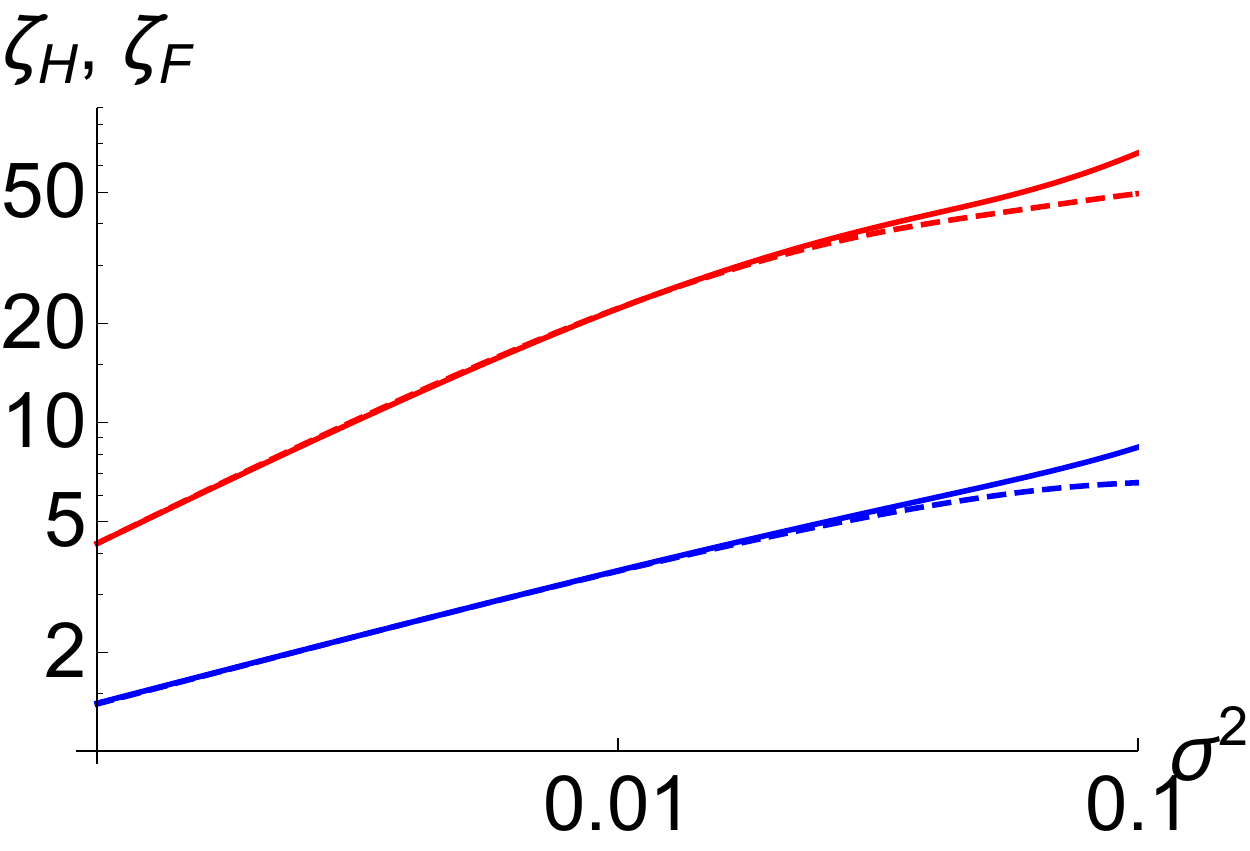}}
\caption{Ratios between QFIs $\zeta_{H}$ (soldi lines) and between FIs $\zeta_{F}$ (dashed lines) corresponding to the two different scenarios as a function of $\alpha$ with fixed $\sigma^2=0.1$ (left panel), and as a function of $\sigma$ with $\alpha=4$ (right panel), for different values of the squeezing parameter $r$: blue lines (bottom), $r=0.5$; red lines (top), $r=1$.}
\label{f:zetaHzetaF}
\end{figure} 

To better quantify and visualize this enhancement we have plotted in Fig. \ref{f:zetaHzetaF} the behaviour of the ratio between the QFIs corresponding of the two scenarios
\begin{align}
\zeta_H = \frac{H_{(a)}}{H_{(b)}} \,,
\end{align}
as before, as a function of both the coherent state amplitude $\alpha$ and the noise $\sigma$, for different values of $r$. We observe that the ratio is always larger than one, implying that the scenario (b) is always more favourable, and more importantly it is also monotonically increasing, saturating towards a fixed value by increasing both $\alpha$ and $\sigma$. 

To better understand the results obtained we have studied the Wigner function \cite{WIGNER1,WIGNER2} of the two families of input states for the two different scenarios. As both states $\varrho_{(a)}$ and $\varrho_{(b)}$ are evidently a Gaussian mixture of Gaussian states, the corresponding Wigner function can be easily evaluated, being in fact a Gaussian mixtures of Gaussian Wigner functions. 
\begin{figure}[tb!]
\centering
\includegraphics[width=0.48\columnwidth, keepaspectratio]{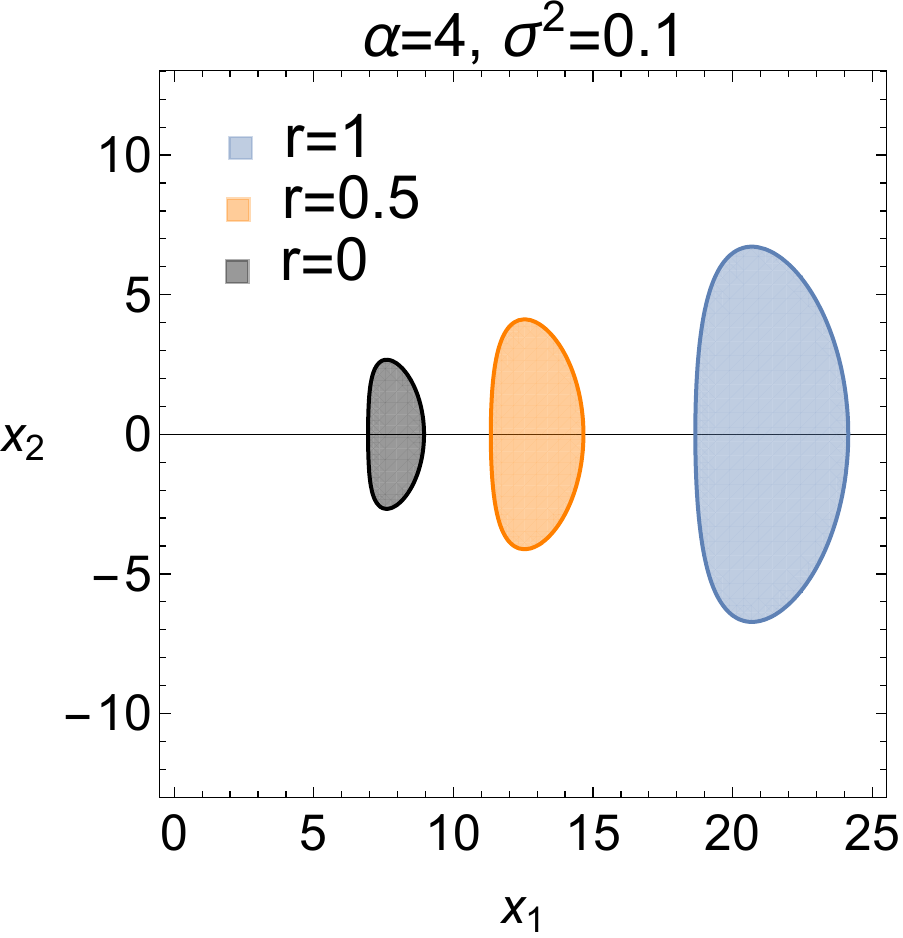}
\includegraphics[width=0.48\columnwidth, keepaspectratio]{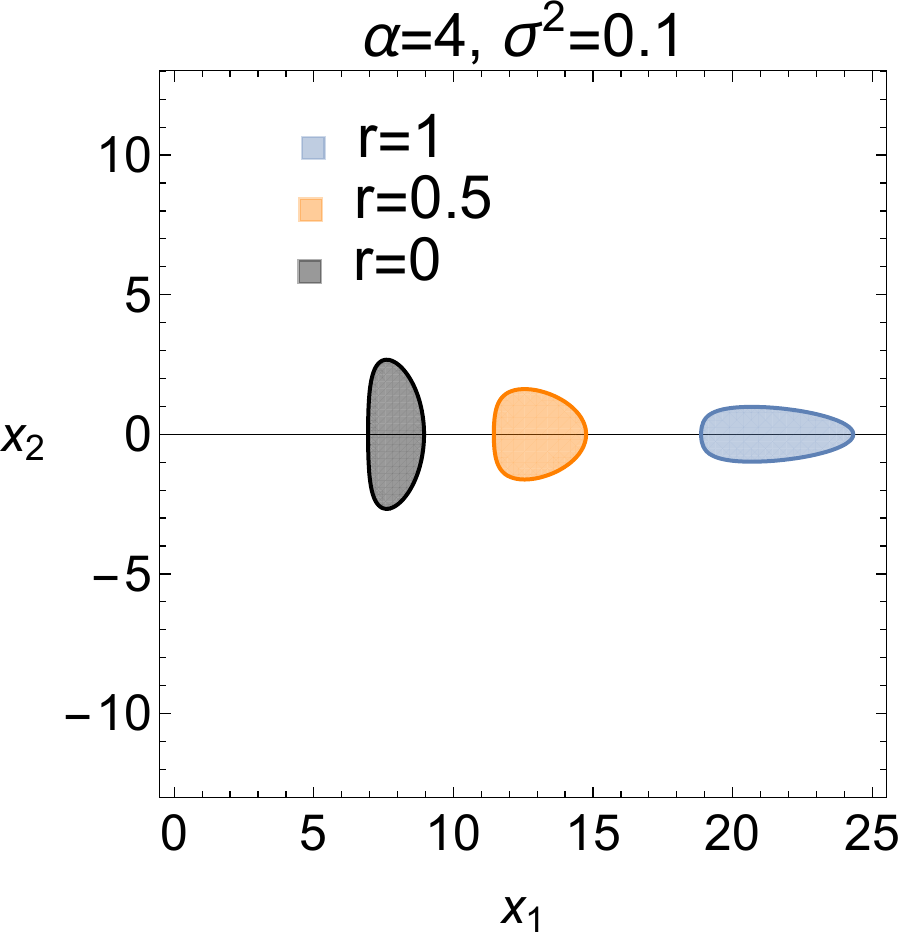}
\caption{Phase space representation of the Wigner function of phase diffused states for both families of input states $\varrho_{(a)}$ (left panel) and $\varrho_{(b)}$ (right panel), for fixed $\alpha=4$ and $\sigma^2=0.1$ and different values of the squeezing parameter $r$: black (left), $r=0$; orange (middle), $r=0.5$; light blue (right), $r=1$.}
\label{ps}
\end{figure}
We also remind that the QFI, as it is clear from Eq.~(\ref{eq:QFIFidelity}), can also be interpreted as a measure of distinguishability of states characterized by an infinitesimal variation of the parameter to be estimated (in our case a phase rotation in phase space). In this sense the phase-space shape of the Wigner function will help in the interpretation of the results just discussed. For example as regards the scenario (a) in Fig.~\ref{f:scenarios}, we observe that the phase noise, occurring after the squeezing operation, completely covers the beneficial effects of  squeezing and the enhancement respect to the state with no squeezing is only due to the increase of the amplitude quadrature $\hat{X}$ mean value, an effect that can be in fact obtained by simply increasing the coherent state amplitude. On the other hand, if the squeezing operation is performed after the noise, as in the scenario (b) in Fig.~\ref{f:scenarios}, one is may counteract the noise by ``squeezing'' the degraded state  and preparing an input state that is clearly more suitable for the purposes of phase estimation: not only the amplitude quadrature $\hat{X}$ has been increased, but, more importantly, the variance of the phase quadrature $\hat{Y}$ is reduced. Remarkably, concerning this last scenario, the possibility to reduce phase diffusion by using an optical parametric oscillator has been recently demonstrated in \cite{CialdiPRL}, thus opening the way to further experimental investigation of this kind of communication scheme with the coherent seed state affected by some phase noise.

\subsection{Performance of homodyne detection}\label{ss:discussion}
We can now discuss the performance of homodyne detection of the $\hat{Y}$ quadrature in the noisy scenario, by assuming that the value of the parameter to be estimated is small, i.e. $\theta \approx 0$. Of course, the latter assumption requires some prior information on the parameter to be estimated, but this can be retrieved by exploiting well-established adaptive estimation strategies. Therefore, here we are dealing with {\em local} estimation theory. Furthermore, if $\theta \approx 0$ we also know that the measurement of the $\hat{Y}$ quadrature yields the largest Fisher information among the homodyne measurements of all the possible quadrature operators.

\begin{figure}[!tb]
\centering
{\includegraphics[width=0.48\columnwidth, keepaspectratio]{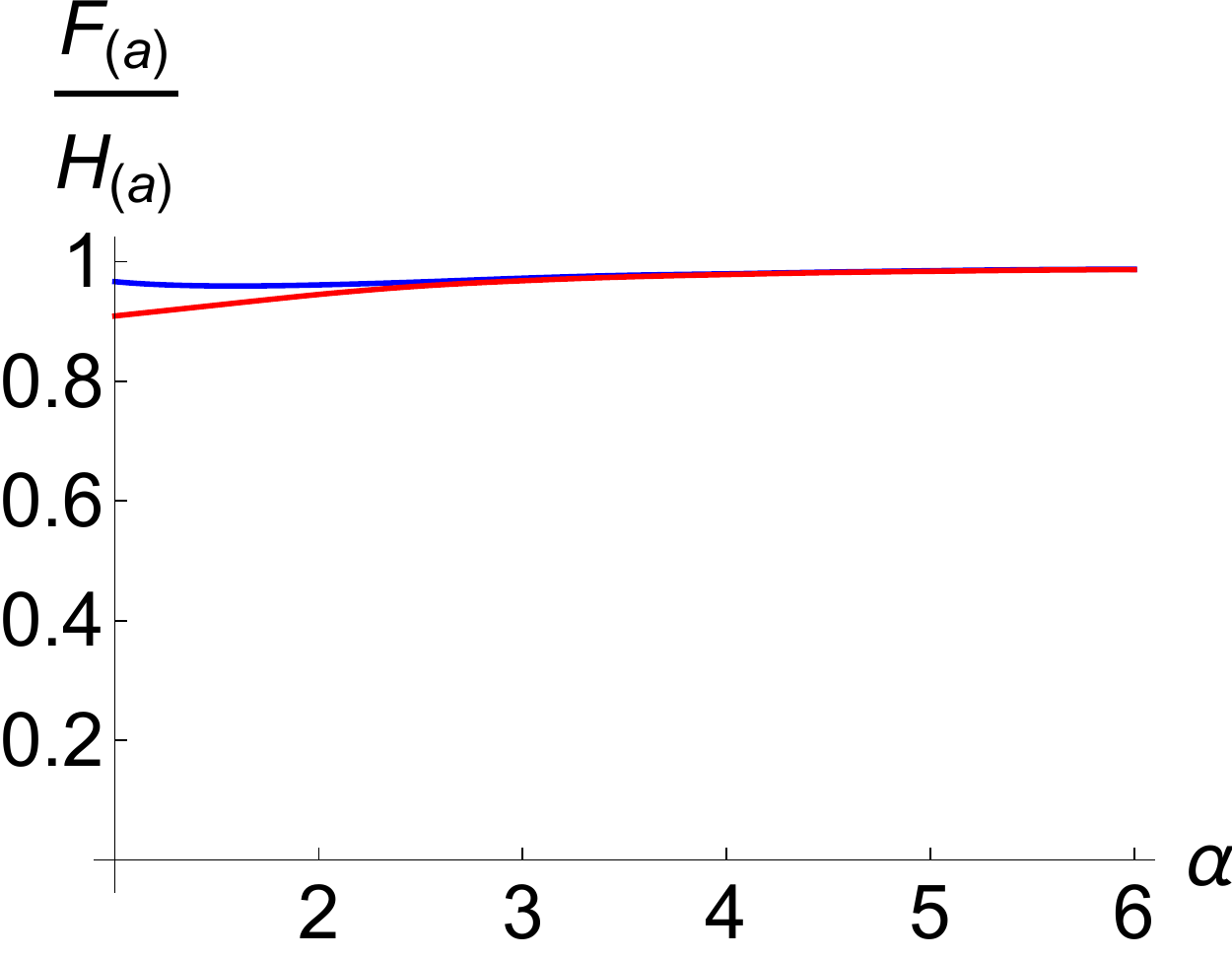}}
{\includegraphics[width=0.48\columnwidth, keepaspectratio]{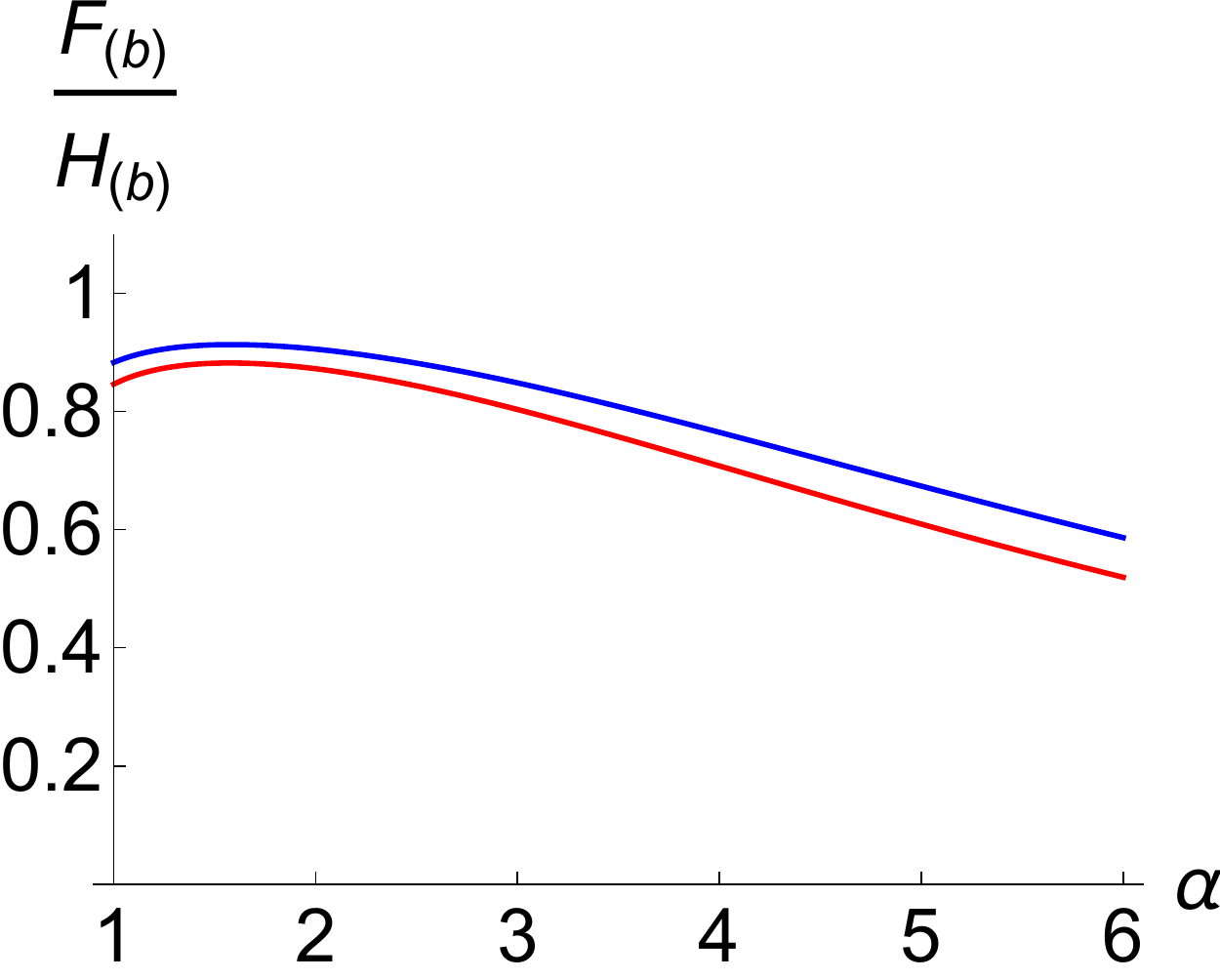}}
\caption{Optimality ratios for the first scenario $F_{(a)}/H_{(a)}$ (left panel) and for the second scenario $F_{(b)}/H_{(b)}$ (right panel) as a function of $\alpha$ ,with fixed noise $\sigma^2=0.1$, and for different values of $r$: blue line (top), $r=0.5$; red line (bottom), $r=1$. }
\label{f:optimalratio}
\end{figure}

As we mentioned above one can easily evaluate the quantum states $\varrho_{(a)}$ and $\varrho_{(b)}$ Wigner functions, and consequently the homodyne probability distribution $p(y|\theta)$ is directly obtained as the marginal of the Wigner functions. The corresponding Fisher informations, denoted as $F_{(a)}$ and $F_{(b)}$ are plotted in Fig. \ref{QFIab_alpha_sigma} as dashed lines. We can immediately observe how in the first scenario homodyne detection of the $\hat{Y}$ quadrature is approximately optimal for all the regimes considered, while this is not the case for the scenario (b), in particular for large phase noise and large coherent amplitude the measurement becomes increasingly less optimal, thus not allowing us to attain the ultimate quantum limit. This behaviour is better represented in Fig. \ref{f:optimalratio} where we have plotted the ratios between FI and QFI for the two scenarios. As we can observe the ratio $F_{(a)}/H_{(a)}$ is approximately equal to one for all the values of $\alpha$, while the ratio $F_{(b)}/H_{(b)}$ shows a not monotonous behaviour and in particular after having reached a maximum around $0.9$, seems to decrease towards zero by increasing the coherent state amplitude. This clearly leaves open the quest for an optimal measurement in the most favourable scenario (b).

On the other hand, we can better compare the performance of homodyne detection in the two scenarios, by introducing the ratio
\begin{align}
\zeta_F = \frac{F_{(a)}}{F_{(b)}} \,.
\end{align}
It is plotted as dashed lines in Fig. \ref{f:zetaHzetaF} as a function of either $\alpha$ and $\sigma$ for different values of $r$. As we expected we find $\zeta_F \le \zeta_H$, but large values $\zeta_F \gg 1$ are still observed showing how the enhancement in the estimation precision obtained by choosing the scenario (b) is still maintained also when we restrict ourselves to measure the output state via homodyne detection.
\par
{Concerning the possible implementation of our estimation scheme in a quantum optical 
system, one should take into account that in  both scenarios the amplitude of the 
coherent state $\alpha$ and the squeezing parameter $r$ are chosen real. This implies that the phase of the squeezing operation, likely to be implemented via an optical parametric 
oscillator, has to be correctly estimated, in order to lock it with the initial phase 
of the coherent state and to the homodyne local oscillator. A complete analysis of these 
aspects, and of their consequences on the estimation strategy will be addressed elsewhere.}
\section{Discussion and conclusions}
\label{IV}
Summarizing, we have first considered a phase estimation scheme using Gaussian squeezed displaced state $|\psi(\alpha,r)\rangle$ as probes. This scheme has been extensively analyzed from a theoretical point of view and an analytic expression of the QFI has been found. In particular has been highlighted the possibility of using squeezing as a powerful tool to increase the QFI and thus the maximum achievable precision. Moreover we have verified that in this noiseless scenario the homodyne detection of the phase quadrature is approximately optimal, that is its corresponding FI is approximately equal to the QFI.

Then, to take into account possible sources of noise, a phase-diffusive noise has been introduced in the generation of the state. By considering a squeezing operation as the resource for the generation of the input probe state, we are left with two possible scenarios: in the first scenario the squeezing operation occurs before the phase noise, whereas in the second scenario squeezing occurs after the phase noise. In both scenarios, since the resulting state is non-Gaussian and non-pure, an analytical expression of the QFI and the FI cannot be found, and a numerical evaluation is required.

Therefore an extended numerical evaluation of both the QFI and the FI has been performed. In the first scenario it has been proven that no sensible improvement can be obtained by the addition of the squeezing. In fact, as seen in the phase space representation of the state, the phase noise completely covers the potential improving effects of the squeezing, thus not allowing to use squeezing to improve phase estimation. However, the homodyne detection of the phase quadrature still remains an approximately optimal measurement for all the values of the parameters involved.

The second scenario on the other hand, has been proven to be more interesting and the numerical evaluation has provided remarkable results: in fact a sensible improvement is always obtainable through the addition of the squeezing, and both QFI and FI result to be larger than the first scenario for all the values of the parameters involved. Once again, these results can be interpreted by looking at the phase space representation of the state. The squeezing, occurring after the phase noise, counteracts the detrimental effects and overall improves the performance of the state in the phase estimation scheme. While homodyne detection in this cases ceases to be the optimal measurement, particularly for larger $\alpha$ and larger $\sigma^2$, the enhancement with respect to the first scenario is still remarkable for all the values of parameters investigated.

\acknowledgments
This work has been supported by UniMI, Project No.~PSR2017-DIP-008, by MAECI, Project No.~PGR06314 ``ENYGMA'', and by MIUR (Rita Levi-Montalcini fellowship). The authors acknowledge useful discussions with A. Pullia.

\end{document}